\documentclass[prb,preprint,showpacs]{revtex4}
\usepackage{amsfonts}
\usepackage{amsmath}
\usepackage{amssymb}
\usepackage{graphicx}
\begin{document}
\title{Quantum interference in nanofractals and its optical manifestation}
\author{F.\ Carlier, V.M.\ Akulin}
\affiliation{Laboratoire Aim\'{e} Cotton, B\^{a}t.\ 505, CNRS\ II, Campus d'Orsay,
ORSAY\ CEDEX F-91405, FRANCE}
\date{\today}

\begin{abstract}
We consider quantum interferences of ballistic electrons propagating 
inside fractal structures with nanometric size of their arms.
We use a scaling argument to calculate the density of states of free 
electrons confined in a simple model fractal.
We show how the fractal dimension governs the density of states and optical
properties of fractal structures in the RF-IR region.
We discuss the effect of disorder on the density of states along with the
possibility of experimental observation.
\end{abstract}

\maketitle

\section{Peculiarity of metallic nanofractals}

Ramified structures are widely observed in nature at scales from the
microscopic world up to the human size. They have been
studied in various contexts and in different domains of science: biology,
physics, chemistry, {\emph{etc.} Surface Science is one particular field where
the ramified semi-metal\cite{refLAC,refLyon}, semiconductor\cite{refBorsella}%
, metallic\cite{refPallmer,refShalaevBottet} or dielectric\cite{Markel,refShalaev}
structures may range from the nanometric up to the micrometric
sizes. The mean free path of electrons in metals is usually of the order of $%
10^2$-$10^3$ $nm$ depending on the kinetic energy.
Therefore electrons propagating ballistically in metallic nanostructures may
manifest essentially quantum behavior associated with strong interference of
their De Broglie waves in contrast to the diffusive\cite{refKhmelnitskii}
or hopping\cite{Shklovskii} behavior intensively studied during the last
decades. The combination of quantum ballistic motion and ramified geometry
suggests to consider the interference of electrons in a fractal metallic
structure confining their propagation.

Tree-like structures is a natural example of fractals. Results obtained for
quantum particles moving on tree-like latices\cite{refDerrida}, for the
quantum localization in the framework of sparse random matrix models\cite
{refMirlin} topologically similar to trees, and for quantum systems with
tree-like hierarchy of interactions\cite{Levitov} have revealed a certain
universality associated with such a topology, that persists in different
physical situations. Therefore for tree-like fractals one can also expect a
universality of the quantum properties related to their specific geometry.
Moreover, the key property of fractal structures is the invariance under
certain
scaling transformations. Therefore considering quantum dynamics of electrons
on fractal trees we take advantage of the scaling arguments\cite{refWilson}.
Note that it is equally important to study the properties of ensembles of
isolated or interacting fractals placed together at a surface, since
it is experimentally difficult to address a single nanometric object. Models
of such ensembles might be also of interest for consideration of conductivity
of thin films\cite{refSaintGobain}, heterogeneous catalysis of nanometer larger
silver particles\cite{refCatalyst}, quantum dot networks\cite{refQuantumdots},
and in other domains.

In this paper, we consider the simplest tree-like fractal with identical
length of the branches at each generation and symmetric nodes as a support
of ballistically propagating electrons. We introduce a single
geometrical parameter $a$
which gives the ratio of branch lengths for successive generations.
We shall see that this
parameter is closely related to the fractal dimension of the tree. We show
that the density of the one-electron states manifests a power law dependence
on the momentum near zero energy with the power index being
the fractal dimension. It is consistent with the result\cite{refMath} for
the low momentum asymptotic of Green functions in systems of fractal
dimensionality. Note that this property is typical of fractals since
linear objects of the same size do not have quantum states close to zero
energy according to the Born-Sommerfeld quantization rule.
We demonstrate the macroscopic manifestations of this power law
in optical properties of surfaces covered by the nanometric ramified
structures by calculating the reflectivity in the RF-IR frequency domain.
Finally, with the help of a simple
random matrix approach\cite{refAkulin} we consider the role of
irregularities in fractal structures resulting from the statistical
distribution of branch lengths and nodes asymmetries, that
does not require to allow for the level-level correlations
in the ballistic regime.

We formulate the problem in terms of the Green functions of a particle
propagating along the fractal. We employ the momentum variable which is
natural for consideration of the interference phenomena, whereas the energy
dependence is given by the dispersion law $E=E(p)$ specific for each type of
systems. It allows one to implement the results for any particular
dependence of the particle energy on the momentum which are usually
different for metals and for semiconductors: for a free particle $E=p^2/2m$,
where $m$ is the mass of the particle, whereas for metals $E=v_f|p|$, where $%
v_f$ is the Fermi velocity.
One-particle Green functions are obtained following the standard quantum field
formalism widely developed in various textbooks\cite{refAbrikosov}.
Quantum state density $g(p)$ and several other properties such as linear
dipole response $R(\omega )$ or conductivity $\sigma (\omega )$ at a
frequency $\omega $ can be found with the help of its retarded $\widehat{G}%
^R(E)$ and advanced $\widehat{G}^A(E)$ Green operators via the
relations\cite{refEfetov} 
\begin{equation}
\label{EQ7}
\begin{array}{c}
g(p)=-\frac 1\pi 
{\rm Im~Tr~}\widehat{G}^R(E) \\ R(\omega )= 
{\rm ~Tr~}\widehat{G}^A(E)\widehat{d}\widehat{G}^R\left( E+\omega \right) 
\widehat{d}\widehat{\rho }(p), \\ \sigma (\omega )={\rm ~Tr~}\widehat{G}^A(E)%
\widehat{j}\widehat{G}^R\left( E+\omega \right) \widehat{d}\widehat{\rho }%
(p) 
\end{array}
\end{equation}
where $\widehat{d}$ is the dipole moment operator, $\widehat{j}$ is the
current operator, and $\widehat{\rho }$ is the density matrix. By the
analogy to a photon propagating in a Fabri-Perot resonator, we can take into
account only the coordinate parts $G(p,x,x^{\prime })$ of the Greens
operators at a given energy $E(p)$ ignoring the resonant denominators $%
(E-E(p)+i0)$. The latter can be factored out during the consideration of the
interference phenomena and have to be restored only at the last stage, prior
to substitution to Eqs.(\ref{EQ7}). Note that in the case of ballistic
propagation the coordinate part of the product $G^A(E)G^R\left( E+\omega
\right) $ can be written in a single factor $G(p)$ depending only on the
momentum $p=\omega .p_E^{\prime }$, where $p_E^{\prime }=1/v_f$,
associated with the energy shift $\omega $.
For $g(p)$, ${\rm Im}~R(\omega ),$ ${\rm Re~}\sigma (\omega )$ the
allowance for denominators yields the Dirac $\delta $-functions of energies
which disappears after taking the trace. Therefore these parameters
responsible for the absorption of electro-magnetic radiation can be
calculated directly when we replace${\rm ~}G^AG^R$ in Eqs.(\ref{EQ7}) by
$G(p)$.
The Kramers-Kronig relation then yields the dispersive parts ${\rm Re~}%
R(\omega )$ and ${\rm Im}~\sigma (\omega ).$ In this paper we therefore call
''Green function'' the coordinate part $G(p,x,x^{\prime })$ of $\widehat{G}%
^A(p_f)\widehat{G}^R(p+p_f)$.

For metals the density matrix is given by the Fermi step $\widehat{\rho }%
(p)=v_fn_e\Theta (-p)$ where $n_e$ is the electron state density in metal
near the Fermi surface and the Fermi momentum is
taken as a reference point. The dipole moment operator $\hat d$ in the
momentum representation reads $\widehat{d}=ie\partial /\partial p$ where $e$
is the electron charge (we set $\hbar =1$), whereas the current operator $%
\widehat{j}$ is simply $pe/m$. Therefore Eq.(\ref{EQ7}) takes the form 
\begin{equation}
\label{EQ8}
\begin{array}{c}
g(p)=\frac 1\pi 
{\rm ~Tr~}G(p) \\ {\rm Im}~R(\omega )=v_fe^2n_e\left. \frac \partial
{\partial p}\right| _{p=0}{\rm ~Tr~}G\left( p+\frac \omega {v_f}\right) \\ 
{\rm Re~}\sigma (\omega )=\frac{e^2n_e\omega }{mv_f}{\rm ~Tr~}G\left( \frac
\omega {v_f}\right) 
\end{array}
\end{equation}
where we have taken into account the relation $\partial \Theta (p)/\partial
p=-\delta (p)$. The trace operation now implies only summation over all
closed trajectories in the coordinate space corresponding to a given
momentum in complete analogy with the Fabri-Perot resonator.

\section{The model of fractal}

We model a fractal by three trees with trunks joint in a node at the fractal
center(Fig.\ref{figTree}). Each of the trees starts with a trunk of length
$L$ and is built by recursive attaching 
at each terminations two homothetical branches scaled by a
factor $a$. The homothetical factor $a$ is the main
parameter of the model.\ It governs all geometrical properties and in
particular the fractal dimension which is the main physical parameter. For $%
a>1$ branches are longer at each step, whereas for $a<1$, branches are
smaller as $n$ increases, which is always the case in our consideration as
we shall see. Electrons propagate ballistically along the trunks and
branches until they reach a node where three branches are attached
symmetrically at the angle $2\pi/3$ as shown in Fig.\ref{figTree}. Nodes
scatter the electrons backwards and forward into the attached branches.

\subsection{Nodes model}

The branches joining a node have different length which depends on the index 
$n$ numerating the generation, that is the number of nodes which separates
the branch from the fractal center. Two branches are of the length $L_n=a^nL$
whereas the branch closest to the trunk has the length $L_{n-1}=a^{n-1}L$.
If we stop the development of the tree at a given $n=N$, the last rightmost
branches have a length $L_N=a^NL$ and the total number of such branches is $%
2^N$.

\begin{figure}[h]
{\centering{\includegraphics*[width=0.5\textwidth]{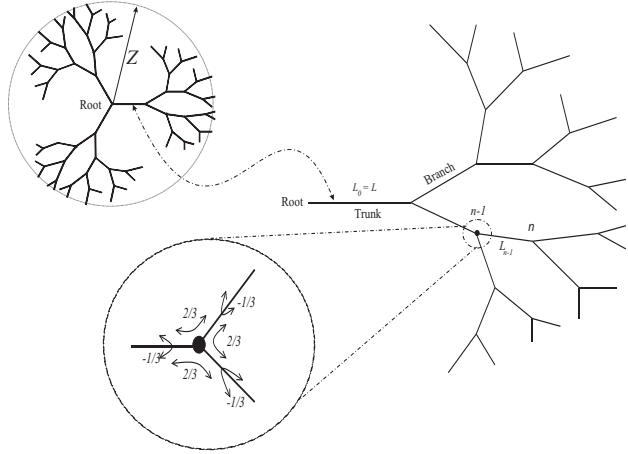}}} 
\caption{Model tree consisting of a trunk of length $L_0=L$ and successive
adjunction of branches. At each generation
$n$, two branches of length $L_n=a.L_{n-1}$ are attached to the previous branch.
The scaling parameter $a$ governs the tree morphological properties, mass, 
length and fractal dimension. We restrict to $1/2<a<1$.
The scattering matrix (Eq.\ref{EQ2}) couples incoming fluxes to outgoing ones.
The fractal of radius $Z$ is built by attaching three identical trees to the root.}
\label{figTree}%
\end{figure}

Having arrived at a node an electron either scatters into the two attached
branches with equal (due to the symmetry) probability or returns back with a
different probability. The node is formally described by a unitary $3\times
3 $ scattering matrix $\hat S$ with the matrix elements $s_{j,j^{\prime }}$
coupling three outgoing probability amplitudes $\phi _j$ of the electron to
the three incoming $\psi _j$ ones, where the marker $j$ assumes the values, $%
l$, $r$, and $b$ for the left-scattered, right-scattered, and the
back-scattered amplitudes, respectively. The relation among the amplitudes
reads

\begin{equation}
\label{EQ1}\left( 
\begin{array}{c}
\phi _b \\ 
\phi _r \\ 
\phi _l 
\end{array}
\right) =\left( 
\begin{array}{ccc}
s_{bb} & s_{br} & s_{bl} \\ 
s_{rb} & s_{rr} & s_{rl} \\ 
s_{lb} & s_{lr} & s_{ll} 
\end{array}
\right) \left( 
\begin{array}{c}
\psi _b \\ 
\psi _r \\ 
\psi _l 
\end{array}
\right) 
\end{equation}

Apart of the unitarity, the matrix $\hat S$ should satisfy two more
requirements imposed by the node symmetry and by the long wave limit. The
symmetry requirement implies that the probability amplitudes for the left-
and the right-scattering given by the coefficients $s_{rb}$ and $s_{lb}$
respectively, are equal. Moreover, the symmetry with respect of the node
rotation at the angle $2\pi /3$ implies that all other off-diagonal
coefficients also have the same value. We also assume that no quantum defect
is associated with the scattering at the node. In the long wave limit it
implies that no phase shift is introduced during the scattering process, and
hence all the parameters $s_{j,j^{\prime }}$ are real. These three
requirements together yield

\begin{equation}
\label{EQ2}\hat S=\left( 
\begin{array}{ccc}
-\frac 13 & ~\frac 23 & ~\frac 23 \\ 
~\frac 23 & -\frac 13 & ~\frac 23 \\ 
~\frac 23 & ~\frac 23 & -\frac 13 
\end{array}
\right) 
\end{equation}
as the only choice for the scattering matrix\cite{refTexier}.

\subsection{Scaling factor and the fractal dimension}

Now we relate the typical length $L$ of the system and the scaling factor $a$
with the fractal dimension employing the self similarity aspect of the
problem. In fact, in the general case $L$ is not the only typical length
scale in the problem.\ The homothetical factor $a$ governs most of the
advanced morphological properties of the model tree. The total length $Z_N$
of the tree with truncated branches of $N+1$-th generation reads

\begin{equation}
\label{EQ3}Z_N=\sum\limits_{k=0}^NL_k=\sum\limits_{k=0}^Na^kL=L\frac{%
1-a^{N+1}}{1-a} 
\end{equation}
This expression imposes a first limit on $a$ :\ for $0<a<1$ the length $Z_N$
converges to a finite value 
\begin{equation}
\label{EQ3.4}Z=\frac{L}{1-a} 
\end{equation}
whereas\ for $a\succeq 1$ it diverges. We consider the fractals of a finite
size only. Actually, the radius of a tree is given by a more complicated
expression and should take into account the geometrical arrangement of the
branches with $2\pi/3$ angle between them. The exact calculation for the
diameter gives $D_N=L(2+a)(1-a^{N+1})/(1-a^2)$ which also converges when $%
N\rightarrow+\infty$ for $a<1$ to the value $D=L(2+a)/(1-a^2)$.

The mass $M_N$ of the tree that is the sum of the lengths of all branches is
given as

\begin{equation}
\label{EQ4}M_N=\sum\limits_{k=0}^N2^kL_k=\sum\limits_{k=0}^N2^ka^kL=\frac{%
1-(2a)^{N+1}}{1-2a}L 
\end{equation}
which converges to $M=L/(1-2a)$ for $a<1/2$ and diverge for $a>1/2$. We are
interested in the regime where the mass of the fractal is infinite, and
therefore $a$ ranges from $1/2$ to $1$.

The model fractal has the same fractal dimension as its consisting
trees. The fractal dimension of a tree is given by a standard evaluation
\cite{refMandelbrot} which is now widely used.
It implies
the calculation of the minimum number $N(\varepsilon )$ of disks of diameter $%
\varepsilon $ needed to completely cover the whole tree.
In a fractal structure, gradual
decreasing of $\varepsilon $ reveals new details causing $%
N(\varepsilon )$ to vary non trivially as $\varepsilon ^{-D_h}$ where $D_h$ 
defines the so called Hausdorff-Besicovitch fractal dimension.

Let us implement this definition in our case of tree-like fractal. In order
to find the number $N(\varepsilon )$ of $\varepsilon $-sized disks required
for covering the tree we make use of the scaling arguments. Let us take the
infinite tree and applying to it the homothetical factor $a$. One obtain
another tree which also has the same infinite structure but starts with a
smaller trunk of length $aL$. This $a$-contracted tree can be considered as
an element of the original tree, namely its first generation branch with all
the branches of subsequent generations attached.\ The size of the discs
covering this branch is apparently $a$ times smaller compared to original
discs of the radius $\varepsilon $. When we attach two $a$-contracted trees
to a trunk of length $L$ we recover the original form of our fractal with
branches covered by $2N(\varepsilon )$ discs of radius $a\varepsilon $. One
requires $L/a\varepsilon $ additional discs to cover the trunk. We therefore
obtain the equation

\begin{equation}
\label{EQ5}N(\varepsilon a)=2N(\varepsilon)+\frac{L}{a\varepsilon } 
\end{equation}
determining an asymptotic behavior of $N(\varepsilon)$ for
$\varepsilon\rightarrow0$.

We look for the solution of Eq.(\ref{EQ5}) in the power-law
form $N(\varepsilon)\sim\varepsilon^{-\alpha}$ with $\alpha>1$. It implies
that the second term in the right hand side of
Eq.(\ref{EQ5}) can be omitted with respect to the first term
as $\varepsilon\rightarrow0$, and
we arrive at $(a\varepsilon)^{-\alpha}=2(\varepsilon
)^{-\alpha}+o(\varepsilon^{-\alpha})$. It yields

\begin{equation}
\label{EQ6}\alpha =-\frac{\ln 2}{\ln a} 
\end{equation}
which is the well-known Hausdorff-Besicovitch fractal dimension of a self
similar recursively built fractal\cite{refMandelbrot}. Equation (\ref
{EQ6}) gives fractal dimension greater than $1$
in the case $a>1/2$ corresponding to an infinite mass $M$.
We also restrict ourselves to the case $a<1$ corresponding to a finite
size $Z$ of fractals. In this regime the spectral peculiarities typical of
such structures manifest themselves in the most interesting way.

\begin{figure}[h]
{\centering{\includegraphics*[width=0.5\textwidth]{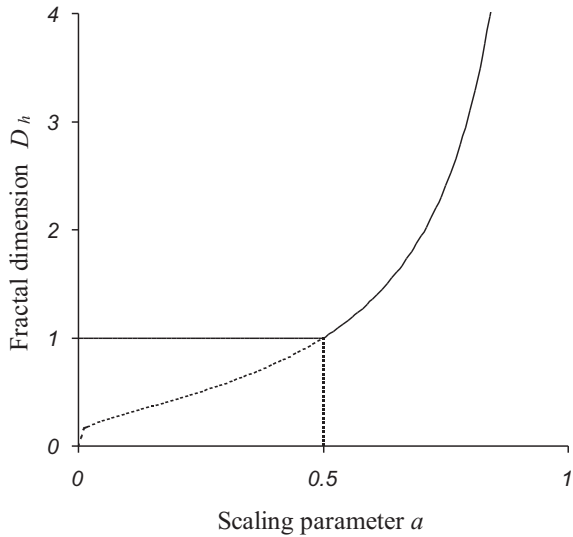}}} 
\caption{Fractal dimension $D_h$ as function of the scaling parameter
$a$. For $0<a<1/2$ the tree has a finite mass and its fractal
dimension is dominated by the trunk : $D_h=1$. For $1/2<a<1$ the mass is infinite
whereas its length remains finite : $D_h=\alpha=-\ln(2)/\ln(a)$}
\label{figDh}
\end{figure}

\section{Green functions and quantization of the fractal states}

The Green functions $G(p)$ generally given by the Feynmann path integral can
be found for the particular case of a tree-like fractal structure from
recurrent relations formulated in terms of the Green function of a free one
dimension particle $G_o(p)=\Theta (x-x^{\prime })\exp [ip(x-x^{\prime})]$
propagating along the branches and the scattering conditions Eq.(\ref{EQ1})
at the nodes. We derive a recurrent relation for these functions and thereby
determine the spectrum of the eigen state density.

\subsection{Recurrent relations}

The idea of derivation of the recurrent relations is illustrated in Fig.\ref
{figMapping}a). By $X_n(p)$ we denote the unknown exact Green function for
the particle leaving a chosen node of $n$-th generation and returning back
after multiple scattering in all the variety of nodes of subsequent
generations connected to the chosen node. Then the Green function $X_{n-1}$
of the previous generation can be considered as a result of the free
propagation of the particle towards the $n$-th node followed by the multiple
scattering at this node resulting in the direct back scattering $s_{bb\text{ 
}}$and the scattering to the attached branches followed by the multiple
returns and back-scattering in the nodes and branches of the subsequent
generations. One finds the result of all these multiple scattering events by
considering the relation Eq.(\ref{EQ1}) among the incoming $\phi $ and
outgoing $\psi $ amplitudes with the allowance for the fact that they are
related by the condition 
\begin{equation}
\label{EQ9}\psi _{l,r}=X_n(p)\phi _{l,r} 
\end{equation}
which holds by the definition of Green functions.

The free propagator $G_o(p)$ gives the relation 
\begin{equation}
\label{EQ10}\widetilde{\psi }=\exp [ipLa^n]\phi _b;\quad \psi _b^{in}=\exp
[ipLa^n]\widetilde{\phi } 
\end{equation}
between the amplitudes $\psi _b$ and $\phi _b$ of the waves incoming to and
outgoing from the node $n$ along the branch attached to the node $n-1$ and
the amplitudes $\widetilde{\phi }$ and $\widetilde{\psi }$ of the waves
outgoing from and incoming to the latter. Here we do not specify whether $%
\widetilde{\psi }^{in,out}$ corresponds to the right scattered or to the
left scattered amplitudes at the node $n-1$ since the relation are identical
for both cases. The scattering matrix Eq.(\ref{EQ2}) and the condition $%
\widetilde{\psi }=X_{n-1}\widetilde{\phi }$ together with Eqs.(\ref{EQ1},\ref
{EQ9},\ref{EQ10}) yield the exact recurrence relation for the Green functions 
\begin{equation}
\label{EQ11}X_{n-1}(p)=\exp [2ipLa^n]\frac{1-3X_n(p)}{X_n(p)-3} 
\end{equation}

Equation (\ref{EQ11}) maps the Green function $X_n(p)$ of a $n$-th node to the
Green function $X_{n-1}(p)$ corresponding to a node of the previous
generation.
As we are interested in the high $n$ behaviour,
this equation has to be inverted to obtain the expression of 
$X_n(p)$ as function of $X_{n-1}(p)$. Changing the $n$ index
to $n+1$ we have

\begin{equation}
\label{EQ11.5}X_{n+1}(p)=\frac{\exp [2ipLa^{n+1}]+3X_n}{3\exp [2ipLa^{n+1}]+X_n(p)}
\end{equation}

that we make use in Fig.\ref{figMapping}b) for $p=0$.

\begin{figure}[h]
{\centering{\includegraphics*[width=0.5\textwidth]{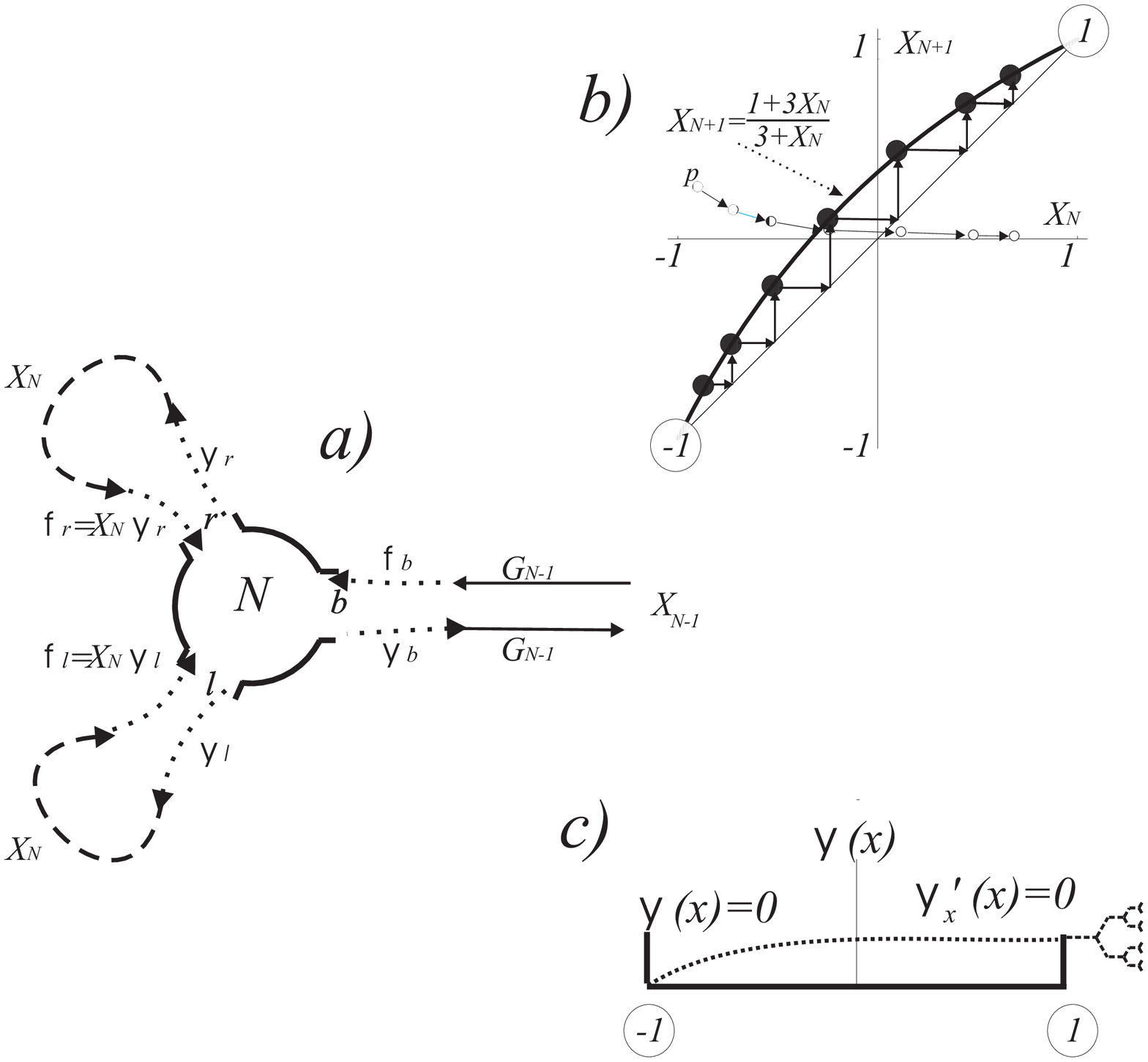}}} 
\caption{Recurrent relations for the Green functions.
(a) A Green function $X_{N-1}$ attached to a parent node consists of a free
propagation $G_0$
followed by the scattering in the node of the next generation and back propagation.
Amplitudes by the outgoing and incoming waves in each branch differ by the
factor $X_{N-1}$. (b) The mapping corresponding to Eq.(\ref{EQ11.5})
in the long wave limit $p=0$
has two stationary points. One corresponds to a regular back-scattering with the phase 
shift $-\pi$, whereas the other does not yield any phase shift and gives rise to
the essentially fractal domain of the spectrum near zero energy. (c) Boundary
conditions at nodes corresponding to the stationary point $-1$ (left), and $+1$ (right).
}
\label{figMapping}
\end{figure}

This mapping Eq.\ref{EQ11.5} has two
stationary points $X_{st}=\mp 1$. Both of them have physical meaning. The
negative sign corresponds to the regular situation when the reflection of
the wave function from a node occurs with a phase shift $-\pi $ exactly in
the same way as the reflection from an infinite vertical barrier implied by
the boundary condition $\psi =0.$ The positive sign corresponds to a free
border $\psi _x^{\prime }=0$ when the wave goes through the node and returns
back with no phase shift, as shown in Fig.\ref{figMapping}c). The latter
case changes the quantization rule for a particle moving in a branch
confined by such nodes from both sides allowing the eigen states at zero
energy that do not exist for the regular confinement. Vicinity of this
stationary point gives rise to a specifically fractal domain of the energy
spectrum at small values of the energies and momenta.

\subsection{Scaling}

Now we make use of the scaling arguments and find the Green function in the
long wave asymptotic and large $n$. The scaling assumption implies that 
$X_{n-1}(p)=X_n(ap)$, which means that the Green functions $X(x)$
corresponding to the branches of any generation are functionally identical
and differ only by scaling of the argument $x=a^np$. Therefore in the long
wave asymptotic where $exp(2IpLa^n)\rightarrow 1$ Eq.(\ref{EQ11}) takes the
form 
\begin{equation}
\label{EQ12}X(x)=\frac{1-3X(ax)}{X(ax)-3} 
\end{equation}
of a functional equation, where we have employed a small dimensionless
argument $x=Lp$ instead of $p$.

This equation has an exact solution 
\begin{equation}
\label{EQ13}X(x)=\frac{1-x^\alpha }{1+x^\alpha } 
\end{equation}
with $\alpha $ given by Eq.(\ref{EQ6}) and yields an asymptotic expression 
\begin{equation}
\label{EQ14}X_n(Lp)=\frac{1-(a^nLp)^\alpha }{1+(a^nLp)^\alpha }. 
\end{equation}

Equation (\ref{EQ14}) holds for small arguments. However, even for a large
values of $x=Lp$ or small $n$ an accurate numerical approximation
can be obtained with the help of few iterations of 
the exact recurrent relation Eq.\ref{EQ11}.
For low $p$ and for $a=2/3$, with Eq.\ref{EQ13} as a starting point,
say $n=10$ iterations of Eq.\ref{EQ11} gives
a good approximation within $1$\%
compared to the exact solution Eq.\ref{EQ14}.

\subsection{Quantization and state density}

Now we are in the position to perform the quantization of the particle
motion on the entire fractal and determine the density of the energy eigen
states. For the purpose we consider the root node at the center of the
fractal with three trunks attached and calculate contributions of all closed
trajectories that start and end in a point of one of these trunks close to
the node. The trajectory sum starts with the zero length trajectory which
gives the contribution $1$. The trajectory first going to the trunk and
returning back gives the contribution $X_0(Lp)$, whereas the contribution of
the trajectory which first goes to the node is $B=(1-3X_0)/(X_0-3)$ according
to Eq.(\ref{EQ11}). The trajectories of the second order give $X_0B$ and $%
BX_0$ whereas the third order results in $X_0BX_0$ and $BX_0B$. The overall
sum reads 
\begin{equation}
\label{EQ15}
\begin{array}{c}
{\rm ~Tr~}G(p)=1+X_0+B+X_0B+BX_0+.. \\ =\frac{(1+X_0)(1+B)}{1-BX_0}=\frac 23%
\frac{1+X_0}{1-X_0}. 
\end{array}
\end{equation}
as it follows from summation of the geometric series.%
\begin{figure}[h]
{\centering{\includegraphics*[width=0.5\textwidth]{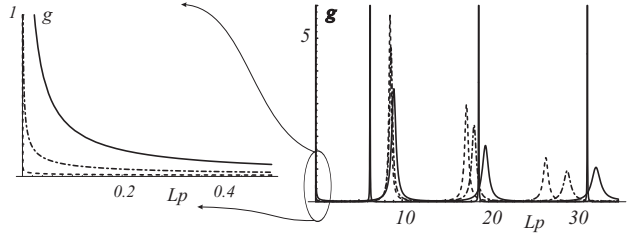}}} 
\caption{Density of states for different fractal dimensions $\alpha=1.18$
corresponding to $a=5/9$ (solid line), $\alpha=1.41$
corresponding to $a=11/18$ (dash-dot line), and $\alpha=1.71$
corresponding to $a=6/9$ (dashed line), calculated with the
help of Eq.(\ref{EQ15}) where $X_0$ has been obtained from Eq.(\ref{EQ14})
after 10 iterations of Eq.(\ref{EQ11}). The fractal diameter
is the same for all fractal dimensions. Vertical lines shows positions
of the levels in a one dimensional potential well of a width equal to the fractal
diameter.}
\label{figSpect}
\end{figure}

In the long wave limit, injecting Eq.(\ref{EQ13}) into Eq.(\ref{EQ15}) we find 
\begin{equation}
\label{EQ16}{\rm ~Tr~}G(p)=\frac 23(Lp)^{-\alpha }, 
\end{equation}
which shows that at small energies the density of fractal energy eigen
states follows the power law dependence on the momentum with the power index
given by the Hausdorff-Besicovitch fractal dimension. In Fig.\ref{figSpect}
we illustrate the difference between the fractal spectrum found from Eq.(\ref
{EQ15}) and the spectrum of a one dimensional particle moving in the
potential well of the width $2Z = 2L/(1-a)$ suggested by Eq.(\ref{EQ3.4})
for the fractal diameter. One clearly sees that the fractal boundary
conditions at the nodes corresponding to the stationary point $X=1$ of
mapping Eq.(\ref{EQ12}) result in the appearance of the spectrum near zero
energy, where the potential well does not have eigenstates.

\section{Nanofractal response to IR-RF field}

Let us consider now the optical response of the nanofractals calculating the
reflectivity of a transparent support surface covered by fractals as a
function of the incident field frequency. We start with the case of isolated
fractals each of which independently contribute to the reflectivity. The
typical frequency domain can be estimated as the inverse of the typical time
of flight of an electron across the fractal given by the Fermi velocity
divided by the fractal size $Z$ Eq.(\ref{EQ3.4}), which for the fractals\cite
{refLAC} of $100nm$ corresponds to the THz frequency domain that is far IR
or short RF radiation. Then we consider the case of ''merging'' fractals,
when the neighboring fractals irregularly placed at the surface can interact
with capacitor-like connections via their most closely approaching
terminations.

\subsection{Isolated nanofractals}

The Maxwell equation 
\begin{equation}
\label{EQ17}\frac{\partial ^2}{\partial x^2}{\bf E}-\frac{\omega ^2}{c^2}%
{\bf E}=\frac{4\pi }{c^2}\delta (\frac xb)\left[ i\omega \sigma _s(\omega )+%
\frac{\omega ^2}cR_s(\omega )\right] {\bf E} 
\end{equation}
for a plane electromagnetic wave ${\bf E}$ incident normally to a surface
covered by isolated fractals at $x=0$ allows one to find an intensity of the
reflected field ${\bf E}_r$ provided the specific conductivity $\sigma
_s(\omega )$ and the specific dipole susceptibility $R_s(\omega )$ of a unit
surface area are known. The inhomogeneities of the surface have to be much
smaller compared to the wavelength of the wave and the thickness $b$ of the
fractal layer. For a wave incident at an angle to the surface the same
equation is valid for the tangent component of the field, whereas the normal
component is not affected by the layer of the fractals. The continuity
condition for the tangent field and the jump of its derivative across the
surface 
\begin{equation}
\label{EQ18}
\begin{array}{c}
{\bf E}+{\bf E}_r={\bf E}_t \\ \frac \omega c\left( {\bf E}-{\bf E}_r-{\bf E}%
_t\right) =\frac{4\pi b}{c^2}\left[ i\omega \sigma _s(\omega )+\frac{\omega
^2}cR_s(\omega )\right] {\bf E}_t 
\end{array}
\end{equation}
yield the relation 
\begin{equation}
\label{EQ19}
\begin{array}{c}
\frac{{\bf E}_r}{{\bf E}}=\frac{-2\pi b\left[ ic\sigma _s(\omega )+\omega
R_s(\omega )\right] }{c^2+2\pi \left[ ic\sigma _s(\omega )+\omega R_s(\omega
)\right] } \\ \simeq \frac{-2\pi b}{c^2}\left[ ic\sigma _s(\omega )+\omega
R_s(\omega )\right] 
\end{array}
\end{equation}
for the ratio of the reflected and the incident field amplitudes.

Equations (\ref{EQ8},\ref{EQ16}) yield 
\begin{equation}
\label{EQ20}
\begin{array}{c}
{\rm Im}~R(\omega )=-N_f\alpha v_f^2\frac{e^2n_e}\omega \frac 23(\frac{%
L\omega }{v_f})^{-\alpha } \\ {\rm Re~}\sigma _s(\omega )=\frac{%
2e^2n_e\omega }{3v_fm}N_f{\rm ~}(\frac{L\omega }{v_f})^{-\alpha } 
\end{array}
\end{equation}
where $N_f$ is the number of fractals per unit area. We replace the product $%
N_fn_e$ by the specific density of states $n_s$ of the fractal material near
the Fermi surface multiplied by the total volume ${\cal V}$ of the material
deposited per unit surface, express the trunk size $L=Z\left( 1-a\right) $
in terms of the typical fractal size $Z$ Eq.(\ref{EQ3.4}) and the fractal
dimension $\alpha $ Eq.(\ref{EQ6}), and substitute $\sigma /v_f\lambda $
instead of the product $n_se^2$ where $\sigma $ is the residual conductivity%
\cite{refLifPit} and $\lambda $ is the electron mean free path in bulk
metal. In the last replacement we assume that $n_s=N_{el}/p_fv_f$ where $%
N_{el}$ is the density of the metal electrons. We arrive at 
\begin{equation}
\label{EQ21}
\begin{array}{c}
{\rm Im}~R_s(\omega )=-\frac{4\sigma \alpha {\cal V}v_f}{3\omega \lambda
\left( 2^{1/\alpha }-1\right) ^\alpha }(\frac{Z\omega }{v_f})^{-\alpha } \\ 
{\rm Re~}\sigma _s(\omega )=\frac{4\sigma {\cal V}\omega }{3\lambda
mv_f^2\left( 2^{1/\alpha }-1\right) ^\alpha }{\rm ~}(\frac{Z\omega }{v_f}%
)^{-\alpha }. 
\end{array}
\end{equation}
Non-analytical behavior of these dependencies at $\omega =0$ does not allow
one to determine the dispersive parts ${\rm Re}~R_s(\omega )$ and ${\rm Im~}%
\sigma _s(\omega )$ from the Kramers-Kronig relations. However the latter
should be of minor importance provided the transparent material supporting
the fractal at its surface has a refraction index $r$ different from $1$. In
the latter case 
\begin{equation}
\label{EQ22}\frac{{\bf E}_r}{{\bf E}}\simeq \frac{{\cal V}b\sigma }{c\lambda 
}\frac{8i\pi (Z\omega /v_f)^{-\alpha }}{3\left( 2^{1/\alpha }-1\right)
^\alpha }\left[ \frac{\alpha v_f}c-\frac \omega {p_fv_f}{\rm ~}\right] +%
\frac{1-r}{1+r}. 
\end{equation}
The simplest possible way to find the missing parts is to take an analytical
continuation of Eq.(\ref{EQ21}) to the complex plane such that $g(Z\omega
/v_f)\sim \omega ^{-\alpha }$ vanishes at the negative part of the real axis.

\subsection{Ensemble of nanofractals}

When the size of the fractals becomes larger than the inter-fractals
distance, the model of isolated fractals fails, since the dipole
approximation for the response is not any longer valid. In the same time,
allowing for the contribution related to the conductivity we have to take
into account the points of the closest approach of neighboring fractals,
where the potential difference experience large changes. These domains work
as capacitors that assume the main part of the dipole activity of the
system. When the ramified structures are randomly distributed on the surface
but not yet result in the electric current percolation, as it is the case
for the experimental work\cite{refLAC} for instance, the fractal ensembles
conform the\ Dykhne model\cite{refDykhne}. Formulated for a two-phase random
conducting surface with the conductivities $\sigma _1$ and $\sigma _2$
different for different phases this model yields the macroscopic
conductivity $\sigma _{eff}=\sqrt{\sigma _1\sigma _2}$ which immediately
suggests 
\begin{equation}
\label{EQ23}\sigma _{eff}=\left[ \frac{4\sigma {\cal V}\omega }{3\lambda
mv_f^2\left( 2^{1/\alpha }-1\right) ^\alpha }{\rm ~}(\frac{Z\omega }{v_f}%
)^{-\alpha }\frac{i\omega b^2}{Zd}\right] ^{1/2} 
\end{equation}
for the effective conductivity of the fractals covering the surface. Here we
have assumed that the capacitors of a plate size $b$ separated by a mean
shortest inter-fractal distance $d$ are subjected to the potential
difference $Z{\bf E}$ accumulated on the distance of the fractal size $Z$.

Substitution of Eq.(\ref{EQ23}) to Eq.(\ref{EQ19}) yields 
\begin{equation}
\label{EQ24}\frac{{\bf E}_r}{{\bf E}}\simeq i\frac{2\pi \omega b^2}{cv_f}%
\left[ \frac{\sigma {\cal V}}{6\lambda mZd\left( 2^{1/\alpha }-1\right)
^\alpha }{\rm ~}\right] ^{\frac 12}\left( \frac{Z\omega }{v_f}\right)
^{-\frac \alpha 2}+\frac{1-r}{1+r} 
\end{equation}
where we have omitted the real part of the effective conductivity as small
relative to the support contribution.

\subsection{Random fractals}

Thus far we have been considering the model of an ideal fractal with a high
symmetry and an exponential variation of the branch lengths with generation
number. In order to get an idea of how close can be such a model to the
reality we now consider an ensemble of irregularly distorted fractals. The
simplest way to model the random distortion is to treat it as a perturbation
of the fractal Hamiltonian by a random matrix with a given mean square $%
\left\langle V^2\right\rangle $ of the matrix elements. The transformation
rule 
\begin{equation}
\label{EQ25}\widetilde{\widehat{G}}(E)=\widehat{G}(\widetilde{E}(E)) 
\end{equation}
\begin{equation}
\label{EQ26}E=\widetilde{E}+\left\langle V^2\right\rangle {\rm Tr}\widehat{G}%
(\widetilde{E}) 
\end{equation}
suggested by one of the authors\cite{refAkulin} as a simple way to solve the
Pastur\cite{refPastur} equation describing such a perturbation. Eq.(\ref
{EQ25}) relates the ensemble averaged perturbed Green function $\widetilde{%
\widehat{G}}(E)$ with the unperturbed one $\widehat{G}(E)$ depending on a
transformed argument $\widetilde{E}(E)$. The transformation $\widetilde{E}%
(E) $ follows from the solution of a nonlinear algebraic equation (Eq.(\ref
{EQ26})) which allows one to find $\widetilde{E}$ for each $E$ selecting
from many possible solutions the one continuously changing from $-\infty $ to 
$\infty $ for $E$ varying in this interval. By the same replacement of the
argument one can obtain all other linear properties of the randomly
perturbed system.

Comparing Eqs.(\ref{EQ7},\ref{EQ8}) and Eq.(\ref{EQ16}) with the allowance
of the condition $g(E<0)=0$ one finds an expression 
\begin{equation}
\label{EQ27}{\rm ~Tr~}\widehat{G}(E)=-g_0\frac 1{{\rm Im(-1)^{-\alpha }}}(%
\frac{-LE}{v_f})^{-\alpha } 
\end{equation}
consistent with the state density Eq.(\ref{EQ7}) for both the positive and
the negative energies. The constant $g_0$ enters as a cofactor of the other
unknown quantity $\left\langle V^2\right\rangle $ and both factors together
form a single energy parameter $W=g_0\left\langle V^2\right\rangle $
responsible for the strength of the random perturbation. We substitute Eq.(%
\ref{EQ3.4},\ref{EQ6},\ref{EQ27}) to Eq.(\ref{EQ26}) and obtain 
\begin{equation}
\label{EQ28}E=\widetilde{E}+\frac W{{\rm \sin }\pi \alpha }\left[ \frac{%
-Z(1-2^{-1/\alpha })\widetilde{E}}{v_f}\right] ^{-\alpha } 
\end{equation}
One sees that by introducing an energy scaling factor $E={\cal E}F$ with $%
F=\left( Z(1-2^{-1/\alpha })/v_f\right) ^{\alpha /(\alpha +1)}W^{-1/(\alpha
+1)}$ equation (\ref{EQ28}) can be reduced to the form 
\begin{equation}
\label{EQ29}{\cal E}=\widetilde{{\cal E}}+\frac{(-\widetilde{{\cal E}}%
)^{-\alpha }}{{\rm \sin }\pi \alpha } 
\end{equation}
which does not contain parameters other than the fractal dimension.

In order to find the universal dependencies $\widetilde{{\cal E}}({\cal E,}%
\alpha )$ there is no need to solve the equation (\ref{EQ29}). After the
replacement $\widetilde{{\cal E}}=-\kappa e^{i\theta }$, one eliminates $%
\kappa $ employing the fact that ${\cal E}$ is real and finds the dependence 
$\widetilde{{\cal E}}({\cal E})$ in a parametric form 
\begin{equation}
\label{EQ30}
\begin{array}{c}
\widetilde{{\cal E}}\left( \theta \right) =-e^{i\theta }\left( \frac{-\sin
\alpha \theta }{\sin \theta {\rm \sin }\pi \alpha }\right) ^{1/(1+\alpha )}
\\ {\cal E}\left( \theta \right) =-\left( \frac{-\sin \alpha \theta }{\sin
\theta {\rm \sin }\pi \alpha }\right) ^{\frac 1{1+\alpha }}\cos \theta
+\left( \frac{-\sin \alpha \theta }{\sin \theta {\rm \sin }\pi \alpha }%
\right) ^{\frac{-\alpha }{1+\alpha }}\frac{\cos \alpha \theta }{{\rm \sin }%
\pi \alpha }. 
\end{array}
\end{equation}
The imaginary part of $\widetilde{{\cal E}}({\cal E},\alpha )/\pi $ shown in
Fig.\ref{figRandom} 
\begin{figure}[h]
{\centering{\includegraphics*[width=0.45\textwidth]{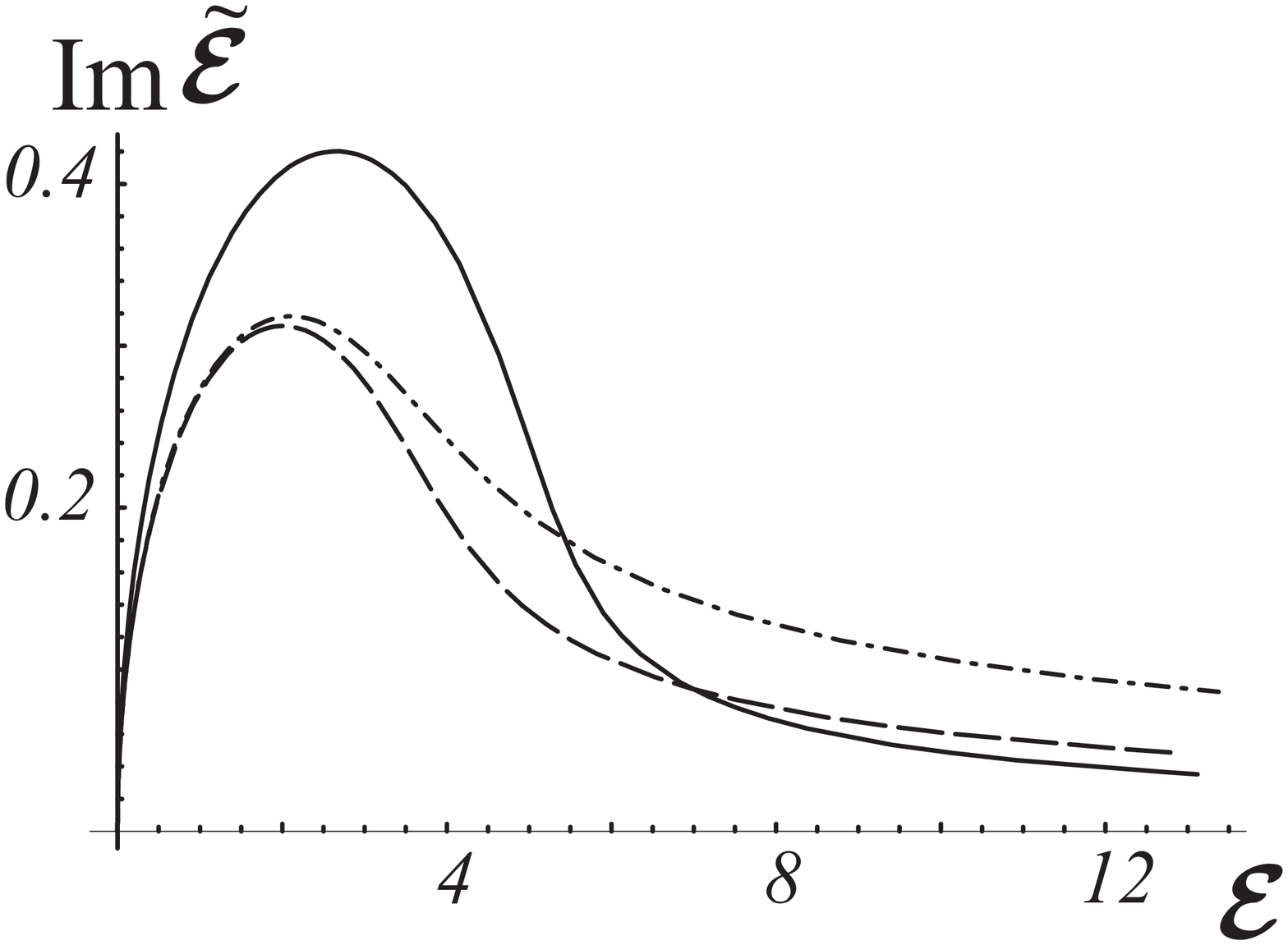}}} 
\caption{Universal forms of the quantum state density profiles for 
randomly perturbed
fractals with the fractal dimensions $\alpha=1.18$ (solid line), $\alpha=1.41$
(dash-dot line), and $\alpha=1.71$ (dashed line). These profiles
do not depend on the size of the perturbation which results only in 
scaling of the energies. }
\label{figRandom}
\end{figure}
as a function of the energy ${\cal E}$ for different fractal dimensions $%
\alpha $ yields the shape of the state density $g_{\alpha ,W}(E)=F{\rm Im}%
\widetilde{{\cal E}}(E/F,\alpha )/\pi $ which for the case of irregular
fractals should replace the factor ${\rm Tr}G(p)/\pi =\frac 2{3\pi }[Z\omega
/v_f(2^{1/\alpha }-1)]^{-\alpha }\sim {\cal E}^{-\alpha }$ in the expression
Eq.(\ref{EQ2}) as well as in Eqs.(\ref{EQ21}) for the dipole response and
the conductivity and in Eq.(\ref{EQ24}) for the effective conductivity of a
disordered surface. It yields 
\begin{equation}
\label{EQ31}{\rm Im}\frac{{\bf E}_r}{{\bf E}}=\left\{ 
\begin{array}{c}
-4\pi ^2 
\frac{{\cal V}b\sigma \omega }{c\lambda }\left[ \frac{v_f}c\frac \partial
{\partial \omega }+\frac 1{p_fv_f}{\rm ~}\right] g_{\alpha ,W}\left( \frac{%
Z\omega /v_f}{2^{1/\alpha }-1}\right) \\ \frac{2\pi \omega b^2}{cv_f}\left[ 
\frac{\pi \sigma {\cal V}g_{\alpha ,W}(E)}{\lambda mZd}{\rm ~}\right]
^{\frac 12}\quad {\rm (merging~fractals)} 
\end{array}
\right. 
\end{equation}
for the absorption of isolated and merging fractals.

\section{Possibility of observation}

We conclude by discussing the possibility to observe the optical
manifestations typical of fractal structures experimentally, for realistic
parameters of nanostructures. We take $p_fv_f\sim 5eV$, $v_f/c\sim 10^{-2}$
for the Fermi velocity and momentum, ${\cal V}\sim 1nm$ for the mean
thickness of the fractal material at the surface, $b\sim 1nm$ for the cross
section size of the fractal branches, $Z\sim \lambda \sim 100nm$ for the
fractal radius of the order of the mean free path on an electron in metal, $%
\sigma [Ag]/\varepsilon _0=6.3~10^7/8.85~10^{-12}\sec ^{-1}$ for the silver
bulk conductivity in CGS units$,$ and $d\sim 10nm$ for the inter fractals
distance. For the frequency $\omega \left[ THz\right] $ we take the units $%
10^{12}Hz$ natural for the electrons moving inside the nanometric sized
objects$.$ In order to be specific we chose the fractal dimension $\alpha
=1.41$ which corresponds to the scaling factor $a=11/18$. In this regime
from Eq.(\ref{EQ31}) one finds 
\begin{equation}
\label{EQ32}{\rm Im}\frac{{\bf E}_r}{{\bf E}}=-10^{-2}\left\{ 
\begin{array}{c}
\left[ 
\frac{\omega \partial }{2\partial \omega }+\omega 10^{-2}\right] g_{\alpha
,W}\left( 0.71~\omega \right) \quad {\rm isolated} \\ 5\omega \left[ {\rm ~}%
g_{\alpha ,W}\left( 0.71~\omega \right) \right] ^{1/2}\qquad {\rm merging} 
\end{array}
\right. 
\end{equation}
which corresponds to the energy absorption at the level of $10^{-4}$. Such a
small absorption is associated however with a phase shift of a few degrees,
which is normally detectable by the ellipsometric measurements in the optical
domain. The same estimate also can serve as the detection limit for IR
domain whereas the internal reflection technique should be even more
sensitive.%
\begin{figure}[h]
{\centering{\includegraphics*[width=0.5\textwidth]{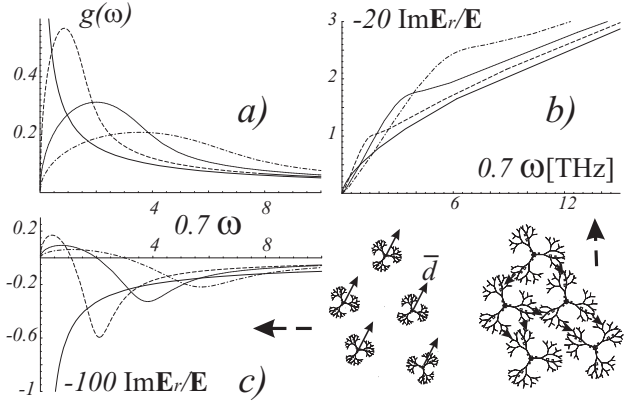}}} 
\caption{Density of states (a), optical response
of surfaces covered by isolated fractals (b) and, merging fractals (c) when
the inter-fractal distance is smaller compared to the fractal size. For
a regular fractal (solid lines) one sees the power law dependencies corresponding to the
fractal dimension  $\alpha=1.41$ chosen. No typical energy reference
exists for the unperturbed fractals, whereas for disordered fractals the typical
energy is given by the disorder parameter $W=g_O <V^2>$ which is small (dashed line) medium
(dotted line) or large (dash-dot line) with respect to the energy unit chosen. }
\label{figManip}
\end{figure}

The dependencies Eq.(\ref{EQ32}) are shown in Fig.\ref{figManip} for
different sizes of the disorder parameter in the regime of both isolated and
merging fractals. The power law dependence corresponding to the ideally
symmetric fractals manifests itself as an asymptotic dependence for the
irregularly perturbed fractals when the frequency exceeds the typical size
of the parameter $W$ governing the disorder.

\section{Acknowledgments}

The authors express their gratitude to D.Khmelnitskii, V.Kravtsov, I.Procaccia and
C.Textier for the discussions and for indication to relevant publications.
One of the authors (V.A) also thanks Ph. Cahuzac and R.Larciprete for the
discussion of the experimental feasibility of ellipsometric and
internal reflection measurements.


\begin{thebibliography}{99}
\bibitem{refLyon}  L. Bardotti, P. Jensen, A. Hoareau, M. Treilleux, and B.
Cabaud, Phys. Rev. Lett., {\bf 74}, 4694 (1995)

\bibitem{refLAC}  C. Br\'echignac, Ph. Cahuzac, F. Carlier, C. Colliex, M.
de Frutos, N. K\'ebaili, J. Le Roux, A. Masson, and B. Yoon, Eur. Phys. J.
D, {\bf 16}, 265 (2001)

\bibitem{refBorsella}  E. Borsella, M.A. Garcia, G.Mattei, C.Maurizio,
P.Mazzoldi, E.Catturuzza, F.Gonella, G.Battaglin, A. Quaranta, and
F.Dacapito, J. of Ap. Phys., {\bf 90}, 4467 (2001)

\bibitem{refPallmer}  S. Pratontep, P. Preece, C. Xirouchaki, R. E. Palmer,
C. F. Sanz-Navarro, S. D. Kenny, and R. Smith, Phys Rev.\ Lett., {\bf 90},
055503 (2003)

\bibitem{refShalaevBottet}  V.M. Shalaev, R. Botet, D.P. Tsai, M. Moskovits,
W.L. Mochan, R.G. Barrera, Physica-A, {\bf 207}, 197 (1994)

\bibitem{Markel}  V.A. Markel, L.S. Muratov, M.I. Stockman and T.F. George,
Phys. Rev. {\bf B 43}, 8183 (1991)

\bibitem{refShalaev}  for a review see V.M.\ Shalaev, Phys.\ Rep., {\bf 272}%
, 61 (1996)

\bibitem{refKhmelnitskii}  Diffusive behavior corresponds both to the
classical conductivity of metallic microfractals and to the multifractal 
structures of the energy eigen functions near
the percolation limit associated with the metal-dielectric transition in the
disordered metals, see D.\ E.\ Khmelnitskii, JETP Lett. {\bf 32} , 229,
(1980); A.\ D.\ Mirlin and F.\ Evers Phys.\ Rev.{\bf \ B\ 62} , 7920 (2000).
It also yields specific optical properties at low frequencies, see U. Sivan
and Y.\ Imry, Phys.\ Rev.\ {\bf B 35}, 6074 (1986)

\bibitem{Shklovskii}  E.\ I.\ Levin, M.\ E.\ Raikh, B.\ I.\ Shklovskii,
Phys.\ Rev.\ {\bf B\ 44} , 11281 (1991)

\bibitem{refDerrida}  B. Derrida and G.J. Rodgers, J. Phys. A, {\bf 26},
L457, (1993)

\bibitem{refMirlin}  A. D. Mirlin, and Y. V. Fyodorov, Phys. Rev. {\bf B 56}%
, 13393 (1997)

\bibitem{Levitov}  B.\ L.\ Altshuler, Y. Gefen, A.\ Kamenev, L.\ S.\
Levitov, Phys. Rev. Lett.,{\bf 78}, 2803 (1997)

\bibitem{refWilson}  see e.g. K.G. Wilson, Phys. Rev. {\bf B 4}, 3174
(1971); K.G. Wilson, Phys. Rev. {\bf B 4}, 3184 (1971)

\bibitem{refSaintGobain}  S. Blacher, F. Brouers, A. Sarychev, A. Ramsamugh
and P. Gadenne, Langmuir, {\bf 12}, 183 (1996)

\bibitem{refCatalyst}  V.I. Bukhtiyarov, A.F. Carley, L.A. Dollard and, M.W.
Roberts, Surf. Sci., {\bf 381}, L605 (1997)

\bibitem{refQuantumdots}  P. Marquardt, Appl. Phys. A, {\bf 68}, 211 (1999)

\bibitem{refMath}  J.M. Barbaroux, J.M. Combes and R. Montcho, J. Math.
Annal. and Appl., {\bf 213}, 698 (1999)

\bibitem{refAkulin}  V.M.\ Akulin, Phys.\ Rev.\ {\bf A48}, 3532 (1993)

\bibitem{refAbrikosov} see e.g. A.A. Abrikosov, L.P. Gorkov, and I.E. Dzyaloshinski,
Methods of quantum field theory in statistical physics, Dover Pub., N.Y. USA (1963); 

\bibitem{refEfetov} K. Efetov, Supersymmetry in disorder and chaos, Cambridge Univ. Press, (1999)

\bibitem{refTexier}  C. Texier and, G. Montambaux, J. Phys. {\bf A 30},
10307 (2001)

\bibitem{refMandelbrot} see for instance J.W. Harris and,H. Stocker,
Handbook of Mathematics and Computational Science, N.Y. Springer-Verlag, (1998)

\bibitem{refPastur}  L.A. Pastur, Theor. Math. Phys. (USSR), {\bf 10}, 67,
(1972)

\bibitem{refLifPit}  E.M. Lifchitz, L.P.\ Pitaevskii Physical Kinetics \S 78

\bibitem{refDykhne} A.M.\ Dykhne, Sov. JETP, {\bf 32}, 63 (1971)
\end{thebibliography}
\end{document}